# Experimental realization of single electron tunneling diode based on vertical graphene two-barrier junction

Rui Xu, Ke-Ke Bai, Jia-Cai Nie, and Lin He*

Department of Physics, Beijing Normal University, 100875 Beijing, People's Republic of China

Usually, graphene is used in its horizontal directions to design novel concept devices. Here, we report a single electron tunneling diode based on quantum tunneling through a vertical graphene two-barrier junction. The junction is formed by positioning a scanning tunnelling microscopy (STM) tip above a graphene nanoribbon that was deposited on a graphite surface. Because of the asymmetry of the two-barrier junction, the electrons can unidirectional transfer from the tip to the graphene nanoribbon but not from the graphene to the tip. This result opens intriguing opportunities for designing new type of graphene transistors in its vertical direction.

Graphene, in which carbon atoms are arranged in a two-dimensional (2D) honeycomb lattice, is considered as a strong candidate for post-silicon electronic devices.[1-7] The unique electronic properties of graphene mainly arise from its gapless, massless, and chiral Dirac spectrum.[8] However, the gapless spectrum (in other words, graphene's metallic conductivity at the Dirac point) blocks the use of graphene in electronic devices because field effect transistors made from graphene remain conducting even when switched off.[9-16] A possible scheme to overcome this problem is to open a band gap in graphene. A potential asymmetry in bilayer graphene,[17,18] quantum confinement,[19] strain,[20-22] and chemical adsorption[23] can result in a gap of graphene and make it semiconducting. Although this scheme works in principle, it also damages the material and degrades graphene's electronic quality. Very recently, Leonid Ponomarenko and his colleagues reported an alternative solution to minimize current leakage by using a new graphene transistor architecture.[24] Their transistor is made of two graphene sheets sandwiched together with an atomically thin insulating barrier. The quantum tunneling of electrons from one layer of graphene to the other can be controlled with an external electric field, i.e., their graphene device can be properly switched on and off.[24] This opens avenue to explore inexhaustibel collection of tunneling transistor to use graphene in its vertical direction.

In this Letter, we report a single electron tunneling diode based on quantum tunneling through a vertical graphene two-barrier junction. The junction is formed by positioning a scanning tunnelling microscopy (STM) tip above a graphene nanoribbon that was deposited on a graphite surface. The atomically thin vacuum between the tip and graphene and between the graphene and the graphite is used as two insulating barriers to control the tunneling of electrons. Because of the asymmetry of the two-barrier junction, the electrons can unidirectional transfer from the tip to the graphene nanoribbon one by one but not from the graphene to the tip.

Samples of highly oriented pyrolytic graphite (HOPG) with the AB Bernal stacking have been chosen to study in our experiments. HOPG samples were cleaved by an adhesive tape in air and transferred into the STM. Upon cleaving a HOPG sample, many different surface structures, such as strained graphene ridge[22] and graphene nanoribbons, can be observed in the resultant graphene layer.[25,26] In this work, only graphene nanoribbons with small area (therefore, with small capacitance) are studied to ensure the observation of single electron tunneling. The STM system was an ultrahigh vacuum four-probe SPM from UNISOKU. All STM and STS measurements were performed at liquid-nitrogen temperature and the images were taken in a constant-current scanning mode. The STM tips were obtained by chemical etching from a wire of Pt(80%) Ir(20%) alloys. Lateral dimensions observed in the STM images were calibrated using a standard graphene lattice. The scanning tunneling spectrum (STS), i.e., the dI/dV-V curve, was carried out with a standard lock-in technique using a 987 Hz a.c. modulation of the bias voltage.

Figure 1(a) shows a typical STM image of a graphene nanoribbon with ~ 3 nm in width and ~ 30 nm in length on HOPG surface. The atomic resolution STM image of the sample is shown in Figure 1(b). The line profile across the nanoribbon reveals that the vertical distance between the graphene nanoribbon and the HOPG surface is about 0.31-0.35 nm, which consists well with the spacing of graphene bilayer. Because of the small area of the nanoribbon and the thin insulating vacuum between the nanoribbon and the graphite, the graphene nanoribbon has a small capacitance $C_S \sim 2.1 \times 10^{-18}$ F, which can be simply estimated by considering two parallel plates separated by 0.34 nm of vacuum. As a consequence, electron hopping between the graphene nanoribbon and the graphite is forbidden at small bias voltages due to the electrostatic energy $e^2/2C_S$ of a single excess electron on the graphene nanoribbon. By positioning a STM tip above the graphene nanoribbon, the three conductive elements, i.e., the STM tip, the graphene nanoribbon, and the HOPG substrate, are isolated from each other by the two insulating vacuum and an asymmetric double-barrier tunnel junction (DBTJ) is formed,[27-32] as shown in Figure 1(c). The physics characters of such system can be described by the transparency rates $\Gamma_S$ ($\Gamma_D$) and the capacitance $C_S$ ($C_D$). The $C_S$ ($\Gamma_S$) and $C_D$ ($\Gamma_D$) are the capacitances (transparency rates) of the nanoribbon-substrate junction and nanoribbon-tip junction, respectively. Taking into account that the transparency rates $\Gamma_S$ ($\Gamma_D$) are mainly determined by the thickness of the insulating barriers, the main features of the DBTJ junction can be captured by an equivalent electronic circuit,[32] as shown in Fig. 1(c). In this classical electronic circuit, the thickness of the insulating barrier is reflected by resistances $R_S$ ($R_D$). The distance between the STM tip and the graphene nanoribbon is about 0.5-1.0 nm (This is a very rough estimate because the absolute tip-height above graphene is not directly measured, which depends on the graphene, the material of tip, and also on the feedback loop parameters.). The asymmetry of the DBTJ junction can results in unidirectional tunnel of electrons from the tip to the graphene nanoribbon.[27-32]

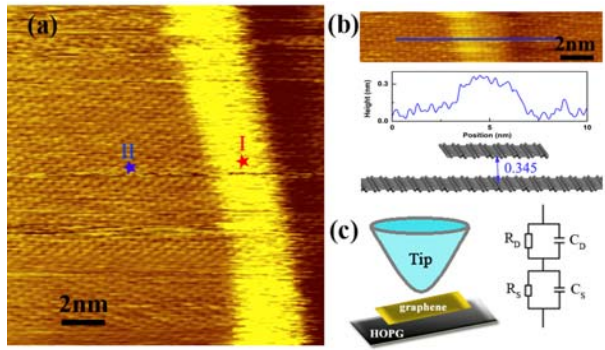

FIG. 1. (Color online) (a) A typical STM image of a graphene nanoribbon on a HOPG surface. The images were taken in a constant-current scanning mode with a tunneling current of 25 pA and a bias voltage of 0.8 V. STS measurements along the nanoribbon, for example, at position labeled by the red star, show a series of peaks corresponding to single electron tunneling for positive bias. The dI/dV-V curve obtained on a flat graphite surface, for example, at position labeled by the blue star, shows featureless V-shaped structure. (b) Top panel: Atomic resolution morphology of the nanoribbon on HOPG surface. Middle panel: Line profile across the nanoribbon (the blue line in the top panel). The vertical distance between the nanoribbon and the HOPG is ~ 0.31-0.35 nm. The width of the nanoribbon is ~ 3.0 nm. Bottom panel: The schematic diagram of the graphene nanoribbon on a graphite surface (only show a topmost graphene sheet for simplicity). (c) The schematic structure model of a STM tip above the graphene nanoribbon that deposited on a graphite surface. The right panel shows the corresponding equivalent electronic circuit of the DBTJ junction.

Figure 2(a) shows a typical tunneling current-voltage (I-V) curve taken at a position, marked by the red star in Fig. 1(a), in the graphene nanoribbon. The unidirectional conductivity of the junction is clearly observed. Several I-V steps corresponding to single electron tunneling is observed for positive bias, while the tunneling current is almost negligible for negative bias. The single electron tunneling diode behavior of the junction is more obviously in the STS measurement, as shown in Fig. 2(b). The spacing of the equidistant peaks is $E_C \sim e/C \sim 0.2$ V, where $C$ is the total capacitance of the whole system. This result indicates that the electrons can unidirectional transfer from the tip to the graphene nanoribbon one by one through quantum tunneling but not from the graphene to the tip. To confirm the observed phenomenon, more than 100 STS spectra are recorded at different positions along the nanoribbon and the main features of these STS spectra are almost reproducible. The spectrum of the graphite surface, which shows the standard featureless V-shaped spectrum (as shown in Fig. 3), is also measured to calibrate the observed behavior. Figure 2(c) shows the theoretical dI/dV-V curve of the DBTJ calculated according to the model developed from the equivalent electronic circuit.[32] The two main features, i.e., the unidirectional conductivity and the single electron tunneling, of the junction are captured by this simple and phenomenological model. The value of $C_S \sim 3.3 \times 10^{-18}$ F is slightly larger than the classical capacitance estimated by two parallel plates, which may arise from interlayer screening, finite thickness of the graphene layer, Fermi velocity reduction and its polarizability.[33]

At a constant bias, the tunneling current represents the distance between the STM tip and the graphene nanoribbon. It decreases with shortening the distance. To study the effect of asymmetry of the two insulating barriers on the performance of the tunneling diode, STS spectra have been recorded at the same position of the nanoribbon with different tunneling currents (different distances), as shown in Fig. 3. By increasing the tunneling current, the asymmetry of the DBTJ weakens and the unidirectional conductivity of the tunneling diode is partially destroyed, as shown in Fig. 3. For the I-V curve measured with the tunneling current ~ 0.84 pA, the maximum ON-OFF switching ratio reaches about ~ 260. This exceeds planar graphene-based field-effect transistors by a factor of 10.[9-16] By replacing one barrier with a high dielectric constant insulator to enhance the asymmetry of the DBTJ, it is expected to achieve a much higher ON-OFF ratio of the single electron tunneling diode in the near future.

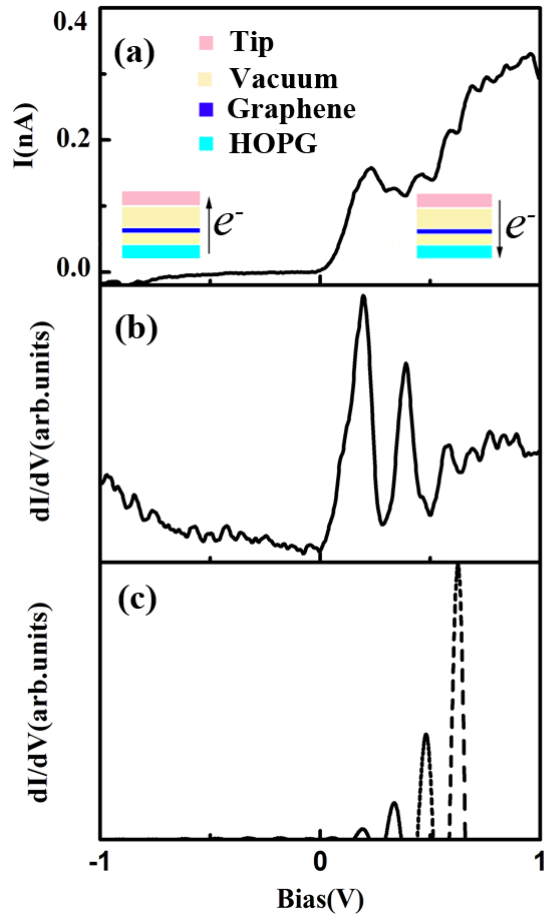

FIG. 2. (Color online) (a) A typical I-V curve taken at a position, marked by the red star in Fig. 1(a), in the graphene nanoribbon. The inset shows schematic structure of our experimental device and the corresponding tunneling process for positive and negative bias. (b) A typical dI/dV-V curve taken taken at the position marked by the red star in Fig. 1(a). (c) The theoretical dI/dV-V curve of a DBTJ calculated according to the model developed from the equivalent electronic circuit, as shown in Fig. 1(c). The corresponding parameters used in the calculation are $R_S = 8 \times 10^8$ Ω, $R_D = 9 \times 10^{10}$ Ω, $C_S = 3.3 \times 10^{-18}$ F (4 μF/cm$^2$), and $C_D = 1.1 \times 10^{-18}$ F (1.3 μF/cm$^2$).

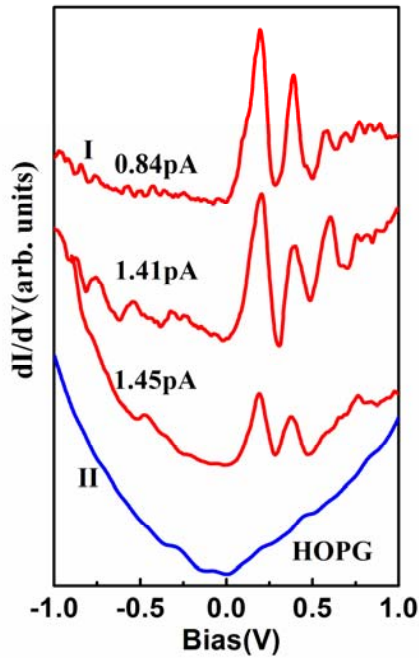

FIG. 3. (Color online) Red curves are STS spectra recorded at the same position, marked by the red star in Fig. 1(a), with different tunnelling currents. The blue curve is a STS spectrum taken on HOPG surface.

In conclusion, we report a single electron tunneling diode based on quantum tunneling through a vertical graphene two-barrier junction. The electrons can unidirectional transfer through the junction one by one due to the asymmetry of the two-barrier junction. The ON-OFF switching ratio of the junction can be easily tuned via changing the asymmetry of the two insulating barriers.

This work was supported by the National Natural Science Foundation of China (Grant Nos. 11004010, 10974019, 51172029 and 91121012) and the Fundamental Research Funds for the Central Universities.

*helin@bnu.edu.cn.

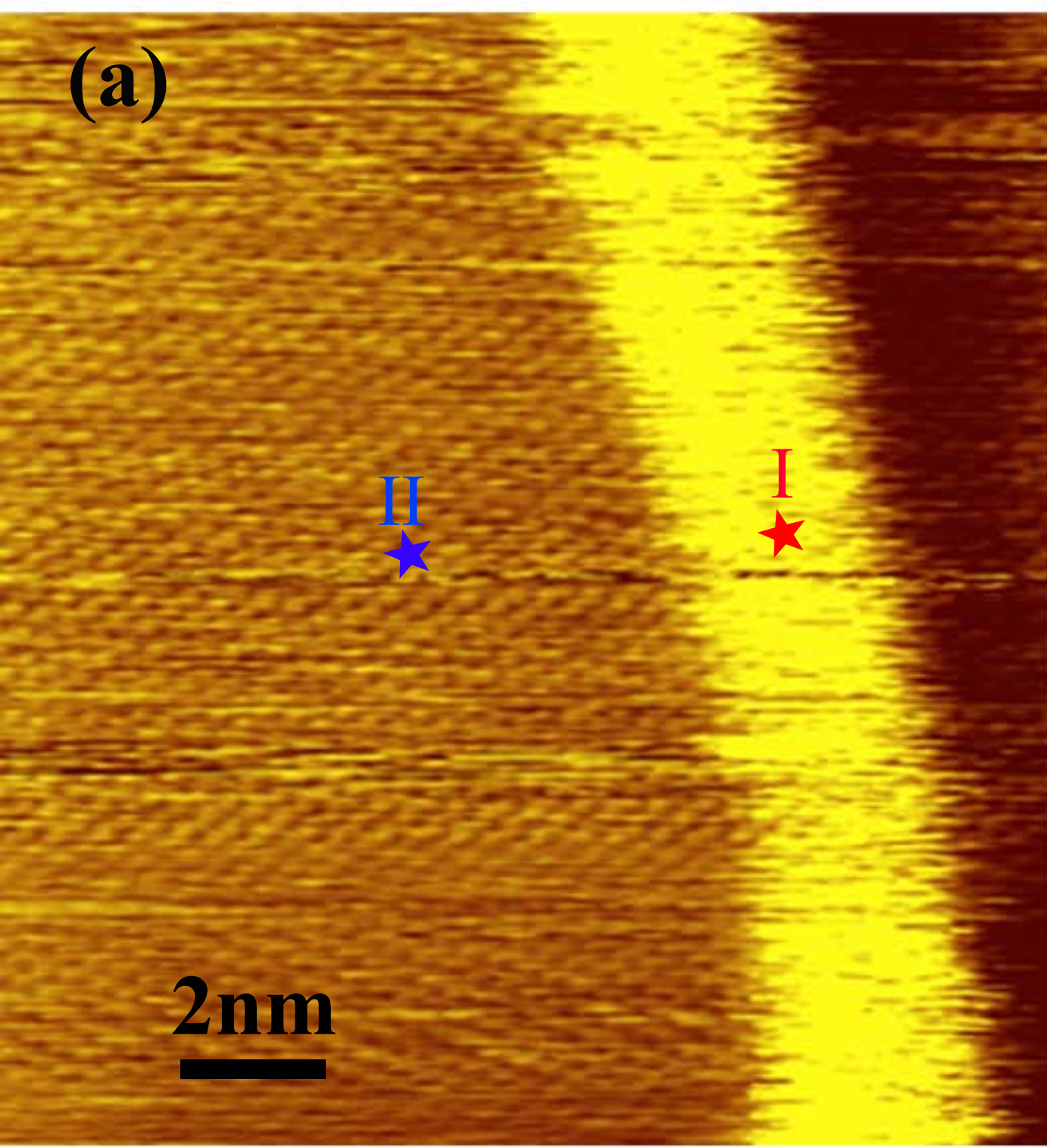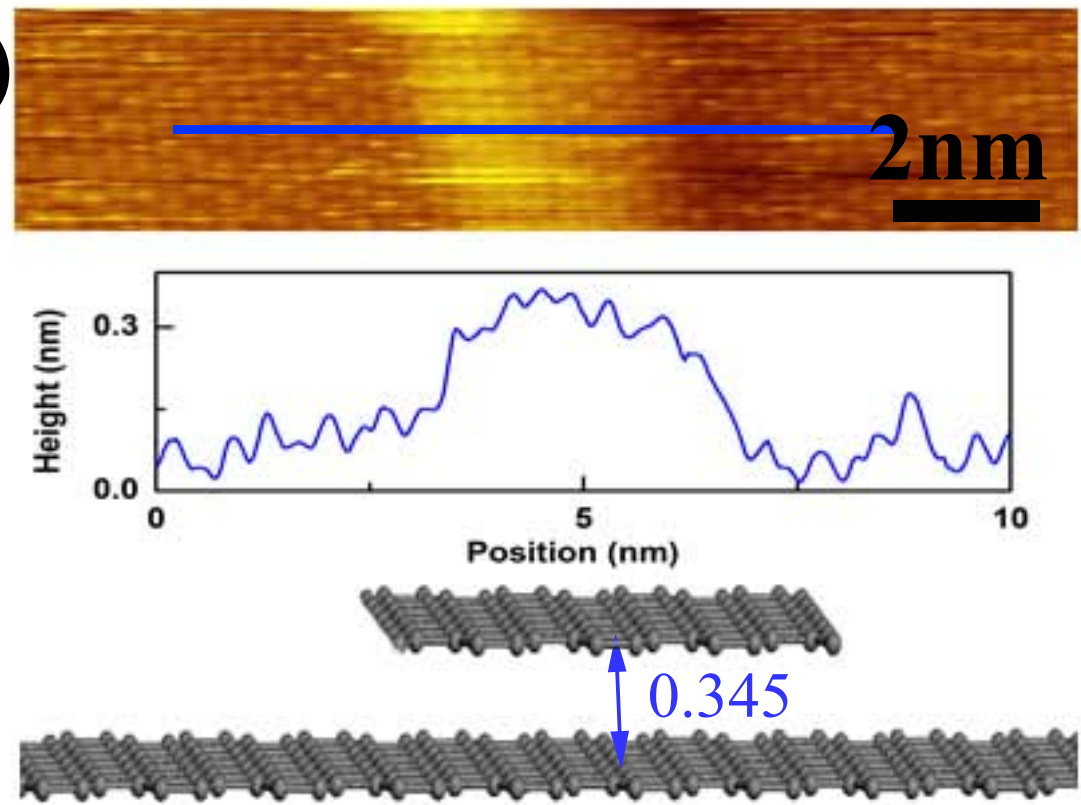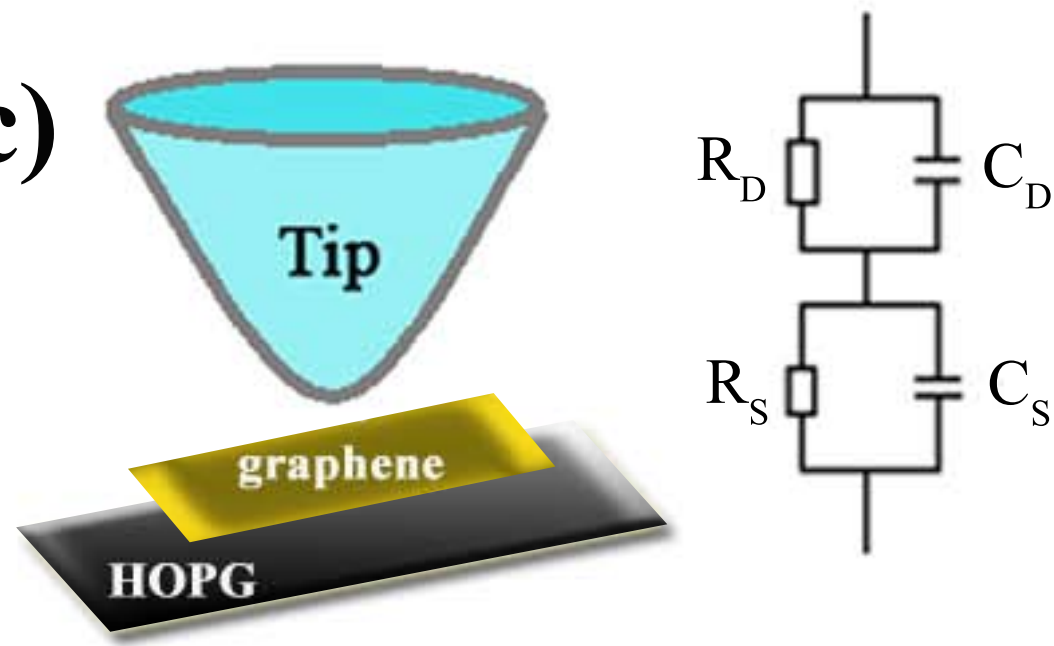

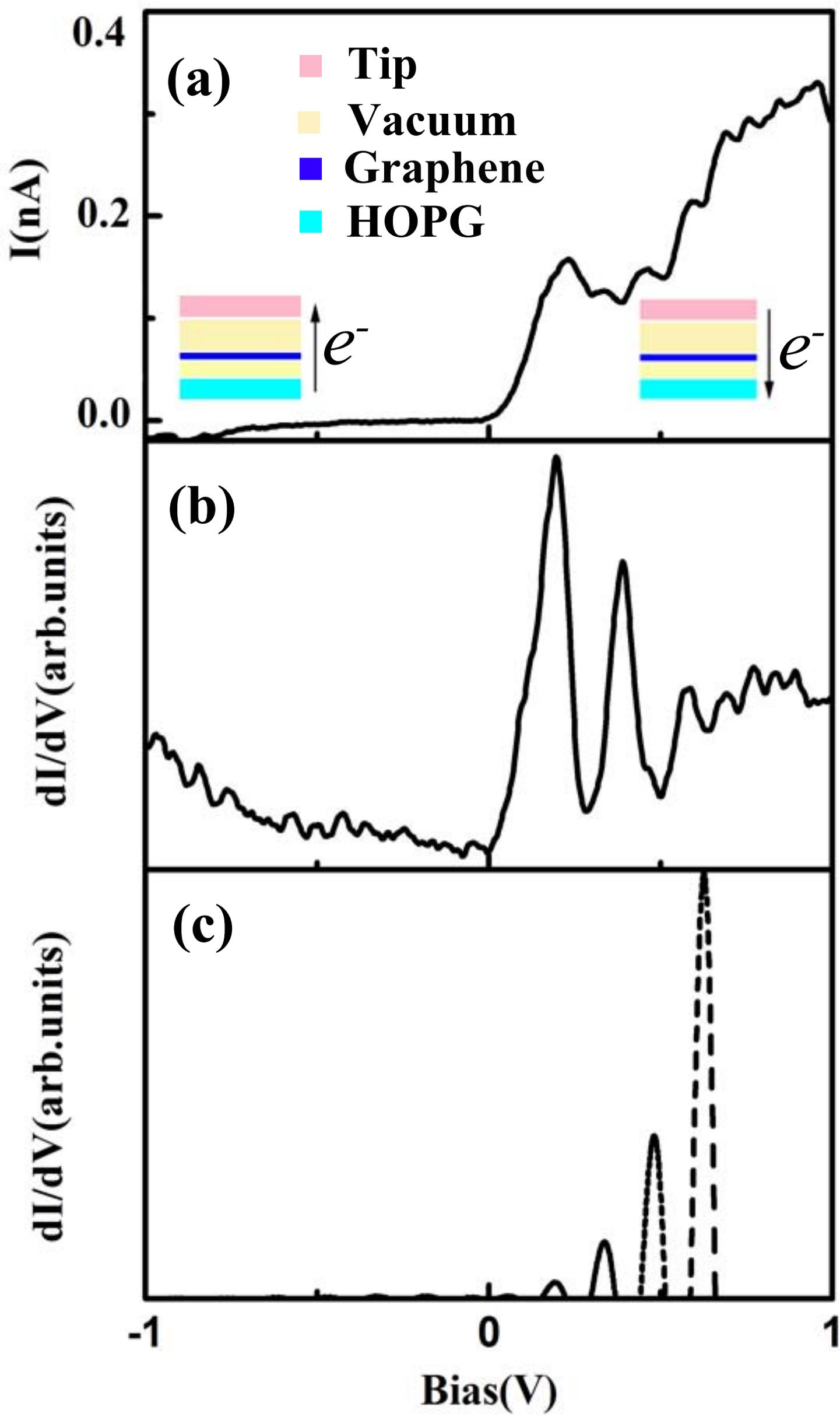

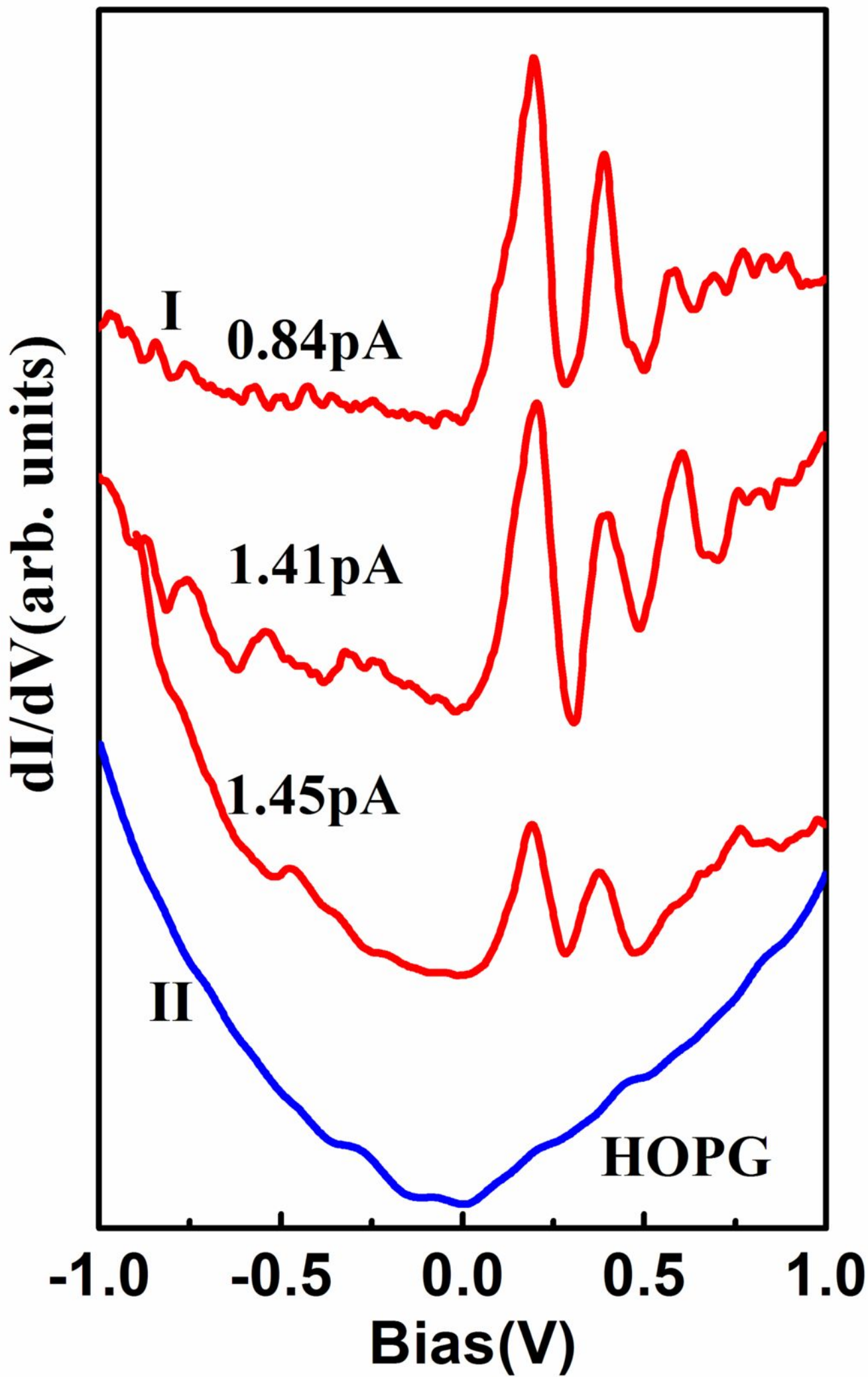